\documentclass[12pt]{article}
\usepackage[fleqn]{amsmath}
\usepackage{amssymb,bm}
\usepackage{cite}
\usepackage{a4}
\usepackage[small]{titlesec}
\numberwithin{equation}{section}

\DeclareMathOperator{\tr}{tr}
\DeclareMathOperator{\Str}{Str}
\DeclareMathOperator{\Tr}{Tr}

\newcommand{\op}[1]{{\hat{\bm{#1}}}}
\newcommand{\MP}[4]{P_{#1}^{(#2)}\!\left(#3;#4\right)}
\newcommand{\hyper}[4]{{}_2F_1
  \biggl(\genfrac{}{}{0pt}{}{#1,\,#2}{#3}\bigg\vert#4\biggr)}
\renewcommand{\Re}{\mathrm{Re}\,}
\renewcommand{\Im}{\mathrm{Im}\,}

\begin{document}
\begin{titlepage}
\vspace*{.5in}
\begin{flushleft}
{\Large\textbf{On the partition function of the six-vertex\\[8pt]
model with domain wall boundary conditions}}
\\[.5in]
\textsc{F. Colomo}
\\[8pt]
\textit{I.N.F.N., Sezione di Firenze\\
and Dipartimento di Fisica, Universit\`a di Firenze\\
Via G. Sansone 1, 50019 Sesto Fiorentino (FI), Italy
}
\\[16pt]
\textsc{A.G. Pronko}
\\[8pt]
\textit{Steklov Institute of Mathematics at St~Petersburg,\\
Fontanka 27, 191023 St~Petersburg, Russia
}
\end{flushleft}
\vspace{1in}
\hrule
\vspace{.2in}
\noindent
\textbf{Abstract}
\bigskip

The six-vertex model on an $N\times N$ square lattice with
domain wall boundary conditions is considered.
A Fredholm determinant representation for the partition
function of the model is given. The kernel of the corresponding
integral operator is of the so-called integrable type, and involves
classical orthogonal polynomials.
{}From this representation, a  ``reconstruction''
formula is proposed, which expresses the partition function
as the trace of a suitably chosen quantum operator, in the spirit of
corner transfer matrix and vertex operator approaches to
integrable spin models.

\vspace{.2in}
\hrule
\vfill
\end{titlepage}
%%%%%%%%%%%%%%%%%%%%%%%%%%%%%%%%%%%%%%%%%%%%%%%%%%%%%%%%%%%%%%%%%%%
\section{Introduction}

The six-vertex model on a square lattice with domain wall
boundary conditions (DWBC) was introduced in \cite{K-82} and
subsequently solved in \cite{I-87}, where a determinant formula for the
partition function was obtained and proven (see also \cite{ICK-92}).
This  model, in its inhomogeneous  formulation ({\sl i.e.} with
the vertex weights given as suitable functions of the position
of the vertex on the lattice), naturally arises when investigating
correlation functions of quantum integrable models in the framework
of quantum inverse scattering method.
In its homogeneous version, the model admits usual interpretation
as a model of statistical mechanics with fixed boundary conditions,
and may be seen as a variation of  the original six-vertex model
with periodic boundary conditions \cite{L-67,L-67a,S-67,LW-72}, which  has been
for decades a paradigmatic  one in  statistical mechanics \cite{B-82}.

The DWBC version of the six-vertex model has however several non trivial
peculiarities which render it worthy of further investigations.
Firstly, the model with this very specific fixed boundary conditions
enjoys  interesting connections with some important issues of enumerative
combinatorics, such as Alternating Sign Matrices
\cite{Ku-96,Z-96,Br-99} and domino tilings \cite{CEP-96,JPS-98}.
Moreover, it appears that, even in the thermodynamic limit, bulk quantities,
such as the bulk free energy, are indeed sensitive to the choice of
boundary conditions \cite{KJ-00,J-00} (see also \cite{BKZ-02}).
Finally, the determinant representation given in \cite{I-87,ICK-92}
for the partition function, and analogous ones recently presented
for the boundary one-point correlation functions (polarizations)
\cite{BPZ-02}, are rather implicit and turn out to be too intricate
for any further, more explicit, answer, except in very particular
cases. Alternative equivalent representations for the
partition function and polarizations would be therefore highly
desirable to address several problems, such as further
Alternating Sign Matrices weighted enumerations, or extensions of the
Arctic
Circle Theorem \cite{JPS-98} beyond the Free-Fermion point.

The model is formulated on a square lattice with arrows on edges. The
only admitted configurations are such that there are always two
arrows pointing away from, and two arrows pointing into, each lattice
vertex; each vertex can therefore be in one out
of six different possible states, a Boltzmann weight $\mathsf{w}_i$
being assigned to each vertex, according
to its state $i$ ($i=1,\dots,6$). We shall consider here the homogeneous
version of the model, where the Boltzmann weights are site independent.
The DWBC are imposed on the $N\times N$ square lattice by fixing
the  direction of  all arrows on the boundaries as follow: the vertical
arrows on the top and bottom
of the lattice point inward, while the horizontal arrows on the left
and right sides point outward.
The correspondence between the Boltzmann vertex weights and the
arrow configurations,  and a typical
configuration of the model with DWBC,  are shown in Fig.~1.

%%%%%%%%%%%%%%%%%%%%%%%%%%%%%%%%%%%%%%%%%%%%%%%%%%%%%%%%%%%%%%%%%%%
%\input{figure1}
\begin{figure}[t]
\unitlength=1mm
% ARROWS
\begin{center}
\begin{picture}(40,40)
% vertex a1
\put(5,30){\line(0,1){10}}
\put(5,37.5){\vector(0,1){1}}
\put(5,32.5){\vector(0,1){1}}
\put(0,35){\line(1,0){10}}
\put(2.5,35){\vector(1,0){1}}
\put(7.5,35){\vector(1,0){1}}
% w1
\put(15,34){$\mathsf{w}_1$}
% vertex a2
\put(35,30){\line(0,1){10}}
\put(35,37.5){\vector(0,-1){1}}
\put(35,32.5){\vector(0,-1){1}}
\put(30,35){\line(1,0){10}}
\put(32.5,35){\vector(-1,0){1}}
\put(37.5,35){\vector(-1,0){1}}
% w2
\put(45,34){$\mathsf{w}_2$}
% vertex b1
\put(5,15){\line(0,1){10}}
\put(5,22.5){\vector(0,-1){1}}
\put(5,17.5){\vector(0,-1){1}}
\put(0,20){\line(1,0){10}}
\put(2.5,20){\vector(1,0){1}}
\put(7.5,20){\vector(1,0){1}}
% w3
\put(15,19){$\mathsf{w}_3$}
% vertex b2
\put(35,15){\line(0,1){10}}
\put(35,22.5){\vector(0,1){1}}
\put(35,17.5){\vector(0,1){1}}
\put(30,20){\line(1,0){10}}
\put(32.5,20){\vector(-1,0){1}}
\put(37.5,20){\vector(-1,0){1}}
% w4
\put(45,19){$\mathsf{w}_4$}
% vertex c1
\put(5,0){\line(0,1){10}}
\put(5,7.5){\vector(0,1){1}}
\put(5,2.5){\vector(0,-1){1}}
\put(0,5){\line(1,0){10}}
\put(2.5,5){\vector(1,0){1}}
\put(7.5,5){\vector(-1,0){1}}
% w5
\put(15,4){$\mathsf{w}_5$}
% vertex c2
\put(35,0){\line(0,1){10}}
\put(35,7.5){\vector(0,-1){1}}
\put(35,2.5){\vector(0,1){1}}
\put(30,5){\line(1,0){10}}
\put(32.5,5){\vector(-1,0){1}}
\put(37.5,5){\vector(1,0){1}}
% w6
\put(45,4){$\mathsf{w}_6$}
\end{picture}
\hskip 1in
%% GRID
\begin{picture}(40,40)
\put(5,10){\line(1,0){30}}
\put(7.5,10){\vector(-1,0){1}}
\put(12.5,10){\vector(-1,0){1}}
\put(17.5,10){\vector(-1,0){1}}
\put(22.5,10){\vector(-1,0){1}}
\put(27.5,10){\vector(-1,0){1}}
\put(32.5,10){\vector(1,0){1}}
\put(5,15){\line(1,0){30}}
\put(7.5,15){\vector(-1,0){1}}
\put(12.5,15){\vector(-1,0){1}}
\put(17.5,15){\vector(-1,0){1}}
\put(22.5,15){\vector(-1,0){1}}
\put(27.5,15){\vector(1,0){1}}
\put(32.5,15){\vector(1,0){1}}
\put(5,20){\line(1,0){30}}
\put(7.5,20){\vector(-1,0){1}}
\put(12.5,20){\vector(-1,0){1}}
\put(17.5,20){\vector(1,0){1}}
\put(22.5,20){\vector(1,0){1}}
\put(27.5,20){\vector(1,0){1}}
\put(32.5,20){\vector(1,0){1}}
\put(5,25){\line(1,0){30}}
\put(7.5,25){\vector(-1,0){1}}
\put(12.5,25){\vector(1,0){1}}
\put(17.5,25){\vector(-1,0){1}}
\put(22.5,25){\vector(1,0){1}}
\put(27.5,25){\vector(1,0){1}}
\put(32.5,25){\vector(1,0){1}}
\put(5,30){\line(1,0){30}}
\put(7.5,30){\vector(-1,0){1}}
\put(12.5,30){\vector(-1,0){1}}
\put(17.5,30){\vector(1,0){1}}
\put(22.5,30){\vector(1,0){1}}
\put(27.5,30){\vector(1,0){1}}
\put(32.5,30){\vector(1,0){1}}
%%%
\put(10,5){\line(0,1){30}}
\put(10,7.5){\vector(0,1){1}}
\put(10,12.5){\vector(0,1){1}}
\put(10,17.5){\vector(0,1){1}}
\put(10,22.5){\vector(0,1){1}}
\put(10,27.5){\vector(0,-1){1}}
\put(10,32.5){\vector(0,-1){1}}
\put(15,5){\line(0,1){30}}
\put(15,7.5){\vector(0,1){1}}
\put(15,12.5){\vector(0,1){1}}
\put(15,17.5){\vector(0,1){1}}
\put(15,22.5){\vector(0,-1){1}}
\put(15,27.5){\vector(0,1){1}}
\put(15,32.5){\vector(0,-1){1}}
\put(20,5){\line(0,1){30}}
\put(20,7.5){\vector(0,1){1}}
\put(20,12.5){\vector(0,1){1}}
\put(20,17.5){\vector(0,1){1}}
\put(20,22.5){\vector(0,1){1}}
\put(20,27.5){\vector(0,-1){1}}
\put(20,32.5){\vector(0,-1){1}}
\put(25,5){\line(0,1){30}}
\put(25,7.5){\vector(0,1){1}}
\put(25,12.5){\vector(0,1){1}}
\put(25,17.5){\vector(0,-1){1}}
\put(25,22.5){\vector(0,-1){1}}
\put(25,27.5){\vector(0,-1){1}}
\put(25,32.5){\vector(0,-1){1}}
\put(30,5){\line(0,1){30}}
\put(30,7.5){\vector(0,1){1}}
\put(30,12.5){\vector(0,-1){1}}
\put(30,17.5){\vector(0,-1){1}}
\put(30,22.5){\vector(0,-1){1}}
\put(30,27.5){\vector(0,-1){1}}
\put(30,32.5){\vector(0,-1){1}}
\end{picture}
\caption{The six allowed types of vertices, their weights and
one of the possible
configurations in the model with the domain wall boundary conditions
for $N=5$.}
\end{center}
\end{figure}
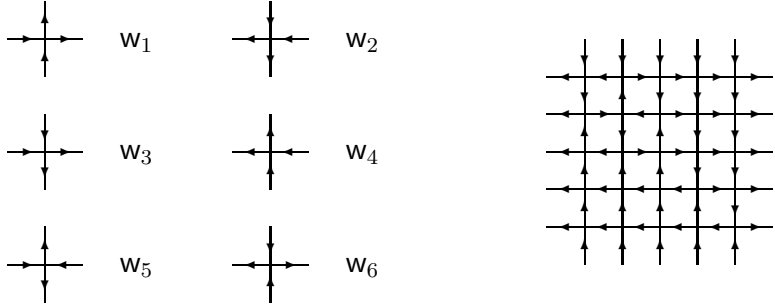
%%%%%%%%%%%%%%%%%%%%%%%%%%%%%%%%%%%%%%%%%%%%%%%%%%%%%%%%%%%%%%%%%%

The partition function is obtained by
summing over all possible arrow configurations, compatible with
the imposed DWBC, each configuration being assigned its Boltzmann weight,
given simply as the product of all the corresponding vertex weights:
\begin{equation}%\label{}
Z_N=\sum_{\text{DWBC configurations}}^{}\
\prod_{i=1}^{6}\mathsf{w}_i^{n_i}\;.
\end{equation}
Here $n_i$ denotes the number of vertices in state $i$, {\sl i.e.}
with Boltzmann weight $\mathsf{w}_i$, in each configuration, and
$\sum_{i=1}^{6}n_i=N^2\;$.
The six-vertex model with DWBC is usually considered with its
weights invariant under inversion of all arrows, and thus with
only  three distinct weight functions,
denoted as
$\mathsf{a}$, $\mathsf{b}$ and $\mathsf{c}$,
\begin{equation}%\label{}
\mathsf{w}_1=\mathsf{w}_2\equiv\mathsf{a}\;,\qquad
\mathsf{w}_3=\mathsf{w}_4\equiv\mathsf{b}\;,\qquad
\mathsf{w}_5=\mathsf{w}_6\equiv\mathsf{c}\;.
\end{equation}
We shall use the following parametrization for the weight functions
\begin{equation} \label{|Delta|<1}
\mathsf{a}=\sin(\lambda+\eta)\;,\qquad
\mathsf{b}=\sin(\lambda-\eta)\;,\qquad
\mathsf{c}=\sin 2\eta\;.
\end{equation}
In terms of this parametrization the result of Ref.~\cite{ICK-92}
for the partition function reads
\begin{align}\label{Z=detZ}
Z_N=\frac{[\sin(\lambda-\eta)\sin(\lambda+\eta)]^{N^2}}
{\prod\limits_{k=1}^{N-1}(k!)^2}\;
{\det}_N H
\end{align}
where $H$ is an $N\times N$ H\"ankel matrix, with entries
\begin{equation}\label{Zmat}
H_{jk}=
\frac{\partial^{j+k}}{\partial\lambda^{j+k}}\,
\frac{\sin2\eta}{\sin(\lambda-\eta)\sin(\lambda+\eta)}\;.
\end{equation}
Here and in the following we use the convention that indices
of $N\times N$ matrices run
over the values $j,k=0,1,\dots,N-1$. Formula \eqref{Z=detZ}
for the partition function, which was originally obtained
within the Quantum Inverse Scattering Method, will be referred
to as the H\"ankel determinant representation.

The purpose of the present paper is to give some other
equivalent representations for the partition function.
The emphasis will
be made on the representation in terms of Fredholm determinant of some
linear integral operator of integrable type, in the sense of paper
\cite{IIKS-90}. Representations of such type have been proven
to be powerful tools in many areas of mathematical physics,
ranging from the theory of Random Matrices
to the asymptotics of Orthogonal Polynomials.
Among them is the theory of correlation
functions of quantum integrable models \cite{KBI-93}, the area of origin
of the six-vertex model with DWBC itself.

The general procedure which we shall follow
to build Fredholm determinant formula for the partition function
has been suggested in \cite{S-00};  in
contrast to that paper,  we shall however apply this procedure
directly to the H\"ankel determinant representation \eqref{Z=detZ}.
In this way, the H\"ankel structure is preserved, and
the integrability of the integral operator
in the Fredholm determinant is ensured by construction.
We therefore propose a simple factorization
for  the determinant of the H\"ankel matrix $H$  appearing in \eqref{Z=detZ}.
This factorization, quite natural in the construction of a
Fredholm determinant representation for the partition function,
allows moreover to identify a core term in the factorized form
of \eqref{Z=detZ}, with all others factor being trivial,
and disappearing with a mere
redefinition of vertex weights. The core term turns out to correspond
exactly to the partition  function for the six-vertex model when
its $R$-matrix is specialized to
the $U_q(\mathfrak{sl}_2)$-invariant $R$-matrix \cite{F-95}, {\sl i.e.}
when its vertex weights are chosen accordingly with its underlying
quantum group symmetry.
This core term, which we shall denote as ${\tilde Z}_N$
gives rise to the Fredholm determinant,
thanks to standard techniques from the theory of Orthogonal Polynomials.

Together with this Fredholm determinant representation, we readily
get for ${\tilde Z}_N$ an equivalent representation  as the
ordinary determinant of an $N\times N$ symmetric matrix, whose
entries can be explicitly evaluated. From this last representation
we propose a ``reconstruction''
formula which expresses the partition function as a trace
of some quantum operator, which, in the large $N$ limit, turns
into the exponential of the boost operator for free fermions
on a lattice. This recalls analogous formulae which appears
in the framework of   corner transfer matrix
and  vertex operator approaches  to
integrable lattice models \cite{B-82,JM-94}.
For $N$ finite, a corresponding ``reconstruction'' formula for
${\tilde Z}_N$ can also be written,  as the trace (over the Fock space
of $N$ canonical fermions) of a product of exponentials of
local operators.

%%%%%%%%%%%%%%%%%%%%%%%%%%%%%%%%%%%%%%%%%%%%%%%%%%%%%%%%%%%%%%%%%
\section{The factorization of the H\"ankel determinant}

In our way to build a Fredholm representation for the partition
function $Z_N$ we shall follow the procedure  suggested
in \cite{S-00}, but applying it
directly to the H\"ankel determinant representation \eqref{Z=detZ}.
As a consequence, the H\"ankel structure is preserved, and
the integrability of the integral operator
in the Fredholm determinant is ensured by construction.
In this section we  shall discuss a specific  factorization
(a somewhat trivial one, but nevertheless basic for what follows)
for  the determinant of the H\"ankel matrix $H$  appearing in \eqref{Z=detZ}.

All the different equivalent representations for
the partition function derived in this paper stem essentially  from
the following simple observation:  using the identity
\begin{equation}%\label{}
\frac{\sin2\eta}{\sin(\lambda-\eta)\sin(\lambda+\eta)}
=\cot(\lambda-\eta)-\cot(\lambda+\eta)
\end{equation}
the matrix $H$ can be naturally written as a difference
of two matrices
\begin{equation}\label{HAA}
H = A_{-} - A_{+}\;,\qquad A_\pm=A\big|_{\phi=\phi_\pm}\;,\qquad
\phi_\pm =\lambda\pm\eta\;,
\end{equation}
where the matrix $A$ can be chosen to be
\begin{equation}\label{Aphi}
A_{jk}
=\frac{\partial^{j+k}}{\partial\phi^{j+k}}
\bigl[\cot\phi-\mathrm{i}\bigr]\;.
\end{equation}
Our choice of the additive constant $-\mathrm{i}$ will
be explained below; its role is to ensure
invertibility of matrix \eqref{Aphi} for any complex value of $\phi$.
The structure of \eqref{HAA} suggests  to
factorize the determinant of $H$, for instance, as follows
\begin{equation}\label{factor}
{\det}_N H=
{\det}_N (A_{-}^{})\;
{\det}_N (I-A_{-}^{-1}\, A_{+}^{})\;.
\end{equation}
The most evident consequence of such  factorization is of course
the relatively  natural and straightforward
emergence of  a   Fredholm determinant representation for the partition
function, the
 Fredholm determinant being related to the second
factor in \eqref{factor}.
The proposed factorization however suggests more: as we shall discuss in
detail in the following (see Sect.~4),  a  ``reconstruction''
formula can be deduced, which allows to represent the partition function
as the trace of a suitably chosen quantum operator, in the spirit of
corner transfer matrix and vertex operator approaches to
integrable spin models.

The analysis of factorization \eqref{factor} relies essentially
on the properties of matrix $A$ and in particular of its determinant.
To this purpose
standard techniques relating H\"ankel matrices with orthogonal
polynomials \cite{S-75} will be exploited.
Let us assume that entries of some generic $N\times N$ H\"ankel
matrix $A$ are
given as
\begin{equation}\label{moments}
A_{jk}:= \int_{-\infty}^{\infty} x^{j+k}
\omega(x)\, \mathrm{d}x\;.
\end{equation}
Let us moreover suppose that there exist a (complete) set of polynomials
$p_n(x)$,
orthonormal with respect to the measure $\omega(x)\,\mathrm{d}x\;$:
\begin{equation}%\label{}
\int_{-\infty}^{\infty}
p_j(x) p_k(x)\,\omega(x)\,\mathrm{d}x
= \delta_{jk}\;.
\end{equation}
Then, calling $\kappa_n$ the leading coefficient of $p_n(x)$,
\begin{equation}%\label{}
p_n(x)=\kappa_n x^n+\dots\;,\qquad \kappa_n \ne 0\;,
\end{equation}
 the determinant of matrix \eqref{moments}
is simply given as
\begin{equation}\label{detA}
{\det}_N A =\prod_{j=0}^{N-1} \kappa_j^{-2}\;.
\end{equation}
Of course, this formula turns out useful provided that the set
of orthogonal polynomials associated to the measure
$\omega(x)\,\mathrm{d}x\;$ can be identified.
In the case
of matrix $H$ given by \eqref{Zmat} appropriate polynomials are
not available and the previous scheme cannot be fulfilled
for generic values of vertex weights.
However, the matrix
defined in equation \eqref{Aphi} is much simpler and
the scheme can be fulfilled explicitly.

The entries of matrix \eqref{Aphi} being periodic
in $\Re\phi$, we may restrict ourselves to consider
values of $\phi$ varying over the vertical strip $0\leqslant\Re\phi<\pi$
(with the point $\phi=0$ excluded). In this region we may use
\begin{equation}\label{cot}
\cot\phi= \text{v.p.}
\int_{-\infty}^{\infty}
\frac{\mathrm{e}^{\phi x}}
{1-\mathrm{e}^{\pi x}}\,\mathrm{d} x
\end{equation}
to write the entries of matrix $A$, Eq. \eqref{Aphi}, in the form
\eqref{moments} with
\begin{equation}\label{weight}
\omega(x)=
\frac{\mathrm{e}^{\phi x}}{1-\mathrm{e}^{\pi x}+\mathrm{i} 0}\;.
\end{equation}
The polynomials $p_n(x)$ (depending on $\phi$ as a parameter),
associated to the matrix $A$, should therefore satisfy
\begin{equation}\label{PfiPfi}
\int_{-\infty}^{\infty}
p_j(x)p_k(x)\,
\frac{\mathrm{e}^{\phi x}}{1-\mathrm{e}^{\pi x}+\mathrm{i}0}\,
\mathrm{d} x
=\delta_{jk}\;.
\end{equation}
To identify the explicit form of these
polynomials we shall now reexpress the orthogonality
condition  in such a way that the integration contour,
though still being the real axis, has no singularity in its vicinity.
This can be achieved by shifting the integration contour
$\mathbb{R} \to \mathbb{R}-\mathrm{i}$ and simultaneously relabelling
the integration variable: $x\to x-\mathrm{i}$. The orthogonality condition
now reads
\begin{equation}\label{PfiPfi2}
\mathrm{e}^{-\mathrm{i}\phi}\int_{-\infty}^{\infty}
p_j(x-\mathrm{i})p_k(x-\mathrm{i})
\frac{\mathrm{e}^{\phi x}}{1+\mathrm{e}^{\pi x}}\, \mathrm{d} x
=\delta_{jk}\;.
\end{equation}
Rewriting the weight function as
\begin{equation}%\label{}
\frac{\mathrm{e}^{\phi x}}{1+\mathrm{e}^{\pi x}} =
\frac{1}{2\pi}\,\Gamma\biggl(\frac{1-\mathrm{i} x}{2}\biggr)\,
\Gamma\biggl(\frac{1+\mathrm{i} x}{2}\biggr)\,\mathrm{e}^{(\phi-\pi/2)x}
\end{equation}
and comparing \eqref{PfiPfi2} with the orthogonality condition of
Meixner-Pollaczek polynomials \cite{KS-98}
\begin{align}\label{MPoc}
&
\frac{1}{2\pi}\int_{-\infty}^{\infty}
\MP{n}{\lambda}{x}{\phi} \MP{m}{\lambda}{x}{\phi}
\Gamma(\lambda-\mathrm{i} x)\,\Gamma(\lambda+\mathrm{i} x)\,
\mathrm{e}^{(2\phi-\pi)x}\,\mathrm{d} x
\notag\\ &\qquad
=\frac{\Gamma(n+2\lambda)}{n!(2\sin\phi)^{2\lambda}}\,\delta_{nm}
\end{align}
where
\begin{equation}\label{MPdef}
\MP{n}{\lambda}{x}{\phi}
=\frac{(2\lambda)_n}{n!}\,\mathrm{e}^{\mathrm{i} n\phi}
\hyper{-n}{\lambda+\mathrm{i}x}{2\lambda}{1-\mathrm{e}^{-2\mathrm{i}\phi}}
\end{equation}
we readily identify the polynomials in question with nothing
but the Meixner-Pollaczek polynomials, where the parameter $\lambda$
(not to be confused with the variable $\lambda$ entering the
parametrization of the vertex weights) must be specialized to
the value $\lambda=1/2$.

Therefore, the polynomials $p_n(x)$ satisfying \eqref{PfiPfi} are
\begin{equation}\label{Pcal}
p_n(x)
=\mathrm{e}^{\mathrm{i}\phi/2}
\sqrt{\sin\phi}\; \MP{n}{1/2}{\frac{x+\mathrm{i}}{2}}{\phi}
\end{equation}
or, due to \eqref{MPdef}, more explicitly
\begin{equation}\label{PcalF}
p_n(x)
=\mathrm{e}^{\mathrm{i}(n+1/2)\phi}
\sqrt{\sin\phi}\;
\hyper{-n}{\mathrm{i} x/2}{1}{1-\mathrm{e}^{-2\mathrm{i}\phi}}\;.
\end{equation}
{}From this expression the highest
coefficient $\kappa_n$ of the polynomial $p_n(x)$
can be explicitly evaluated
\begin{equation}\label{kappa}
\kappa_n(\phi)=\frac{\mathrm{e}^{\mathrm{i}\phi/2}
(\sin\phi)^{n+1/2}}{n!}\;,
\end{equation}
and, due to formula \eqref{detA}, we readily get
for the determinant of the matrix $A\;$:
\begin{equation}\label{detAphi}
{\det}_N A = \frac{\mathrm{e}^{-\mathrm{i} N\phi}}{(\sin\phi)^{N^2}}
\prod_{n=1}^{N-1} (n!)^2\;.
\end{equation}
This last expression shows that if matrix $A$ exists
(that is if all its entries are finite) then it is invertible,
since its determinant never
vanishes (for finite $\phi$).
Factorization \eqref{factor}
therefore leads to an equivalent representation for the partition
function, valid for all  (non vanishing) values of $\phi_\pm$.

Let us note that our choice of the constant
$(-\mathrm{i})$ in the definition of matrix \eqref{Aphi} can be easily
explained now. Indeed,
by considering a combination of \eqref{detAphi} and its formal
complex conjugate ($\mathrm{i}\to\mathrm{-i}$) one finds
\begin{equation}%\label{}
{\det} \left[\frac{\partial^{j+k}}{\partial\phi^{j+k}}
\bigl(\cot\phi+\alpha\bigr)\right]_{j,k=0}^{N-1}=
\frac{\cos(N\phi)+\alpha\sin(N\phi)}{(\sin\phi)^{N^2}}
\prod\limits_{n=1}^{N-1}(n!)^2\;.
\end{equation}
The expression
$\cos(N\phi)+\alpha\sin(N\phi)$ possesses zeroes in the complex plane
of the variable $\phi$ unless $\alpha=\pm\mathrm{i}$. Thus, by choosing
$\alpha=-\mathrm{i}$ (or, equivalently, $\alpha=\mathrm{i}$)
the invertibility of matrix $A$ is ensured.

With \eqref{detAphi} taken
into account the partition function now reads
\begin{equation}\label{Znew}
Z_N = \bigl[\sin\phi_{+}^{}\bigr]^{N^2}
\mathrm{e}^{-\mathrm{i}N\phi_{-}^{}}\;
{\det}_N (I-A_{-}^{-1}\, A_{+}^{})\;.
\end{equation}
It is worth emphasizing that
the partition function is, in fact,
described only by the last factor, the first two
factors having a trivial meaning, and disappearing with a mere redefinition
of the vertex weights.
Indeed, the first factor in \eqref{Znew} can be seen as the result of
a  common prefactor
in all weights, and we get rid of it by changing the
 overall normalization of the weights.
The second one is a ``boundary'' factor
specific of the DWBC choice for the six-vertex model; it can be
removed by introducing a suitable asymmetry in weights
$\mathsf{w}_5$ and $\mathsf{w}_6$, since  any configuration
contributing to the partition function (that is, satisfying both the
``ice rule'' and the DWBC), is such that the numbers of vertices
of these two  types satisfy the condition
$\# \mathsf{w}_6 -\# \mathsf{w}_5=N$,
Hence, by choosing the weights to be
\begin{align}\label{uqsl2}
&\mathsf{w}_1=\mathsf{w}_2=1\;,\qquad
\mathsf{w}_3=\mathsf{w}_4=\mathsf{b}/\mathsf{a}\;,
\notag\\ &
\mathsf{w}_5=\mathrm{e}^{-\mathrm{i}(\lambda-\eta)}\mathsf{c}/\mathsf{a}\;,
\qquad
\mathsf{w}_6=\mathrm{e}^{\mathsf{i}(\lambda-\eta)}\mathsf{c}/\mathsf{a}\;,
\end{align}
with $\mathsf{a}$, $\mathsf{b}$, and $\mathsf{c}$ given
by \eqref{|Delta|<1}, the partition function reduces to the
sole determinant in \eqref{factor}.

The choice of the weights \eqref{uqsl2} has a simple interpretation
in terms of the six-vertex model $R$-matrix, the matrix of local vertex
states. In the standard notations
the choice of the weights in the form \eqref{uqsl2} corresponds to the
following $R$-matrix
\begin{equation}\label{Rmatrix}
R(\nu)=
\begin{pmatrix}
1 &&&\\& \beta(\nu) & \mathrm{e}^{\mathrm{i}\nu}\gamma(\nu) & \\
& \mathrm{e}^{-\mathrm{i}\nu}\gamma(\nu) & \beta(\nu) & \\
&&& 1
\end{pmatrix}
\end{equation}
where
\begin{equation}\label{bgnu}
\beta(\nu)=\frac{\sin\nu}{\sin(\nu+2\eta)}\;,\qquad
\gamma(\nu)=\frac{\sin2\eta}{\sin(\nu+2\eta)}\;,\qquad
\nu=\lambda-\eta\;.
\end{equation}
In this expression one easily recognizes
the $U_q(\mathfrak{sl}_2)$-invariant $R$-matrix \cite{F-95} which is,
moreover, normalized to satisfy the ``unitarity condition'':
\begin{equation}\label{unitarity}
R(\nu)\,\bigl( P R(-\nu) P\bigr)=I\;,
\end{equation}
with the permutation operator $P :=R(0)\,$.
Thus, the factorization \eqref{factor} of the initial H\"ankel determinant
leads naturally to the representation of the
partition function just as the sole determinant in \eqref{Znew},
provided that the vertex weights are chosen according  to the
underlying quantum group symmetry of the model.
{}From now on we shall assume this choice, equation \eqref{uqsl2},
for  the vertex weights, denoting the corresponding partition function
as ${\tilde Z}_N$, given as
\begin{equation}\label{Zuqsl2}
{\tilde Z}_N
={\det}_N (I-A_{-}^{-1}\, A_{+}^{})\;.
\end{equation}

It is worth  mentioning here that almost everything above
can also be extended to the case of the determinant formula of
Ref.~\cite{I-87} for the partition function of the inhomogeneous model.
Without giving any detail, let us just emphasize that
previous considerations become  even more transparent, since
in that (more general) case matrix $A$ is a Cauchy matrix.
However, the homogeneous model possess the further interesting
property that the partition function can be expressed as the
Fredholm determinant of an integral operator of integrable type,
in the sense paper of \cite{IIKS-90}.
This will be the subject of the next Section.

%%%%%%%%%%%%%%%%%%%%%%%%%%%%%%%%%%%%%%%%%%%%%%%%%%%%%%%%%%%%%%%%%
\section{The partition function as a Fredholm determinant}

We shall now focus our attention on the determinant
in \eqref{Zuqsl2}. First of all we need to build
the entries of matrix $A_{-}^{-1}=A^{-1}|_{\phi=\phi_{-}}$.
This task can be achieved straightforwardly by borrowing
standard techniques from the theory of orthogonal
polynomials \cite{S-00,S-75}. Let us recall  that,  indeed,
once the set of orthogonal
polynomials  associated to a given  H\"ankel matrix $A$ with entries
 \eqref{moments} is known,
the entries of the corresponding inverse matrix $A^{-1}$ can
be evaluated from the function
\begin{equation}\label{KN}
\mathcal{K}_N(x,y) =\sum_{n=0}^{N-1} p_n(x) p_n(y)\;.
\end{equation}
simply in terms of  partial derivatives:
\begin{equation}\label{inverse}
A^{-1}_{jk}=\frac{1}{j!}\frac{\partial^j}{\partial x^j}
\frac{1}{k!}\frac{\partial^k}{\partial y^k}\,
\mathcal{K}_N(x,y) \bigg|_{x=0,\; y=0}.
\end{equation}
The proof is based on the fact that function $\mathcal{K}_N(x,y)$
is the kernel of an integral operator,
with respect to the measure $\omega(x)\mathrm{d}x\,$.
This operator, by construction,  projects over the subspace of
polynomials of order less than $N$, and therefore
acts as identity operator
on monomials $1,x,\dots,x^{N-1}\;$:
\begin{equation}\label{Kproj}
\int_{-\infty}^{\infty}
\mathcal{K}_N(x,y)\,y^m\omega(x)\,\mathrm{d}x=x^m\;,\qquad
m=0,1,\dots,N-1\;.
\end{equation}
With \eqref{Kproj} taken into account, equation \eqref{inverse}
can be verified directly:
\begin{align}%\label{}
\sum_{k=0}^{N-1} A^{-1}_{jk} A_{km}^{}
&=\sum_{k=0}^{N-1}
\frac{1}{j!}\frac{\partial^j}{\partial x^j}
\frac{1}{k!}\frac{\partial^k}{\partial y^k}\; \mathcal{K}_N (x,y)
\int_{-\infty}^{\infty}
z^{k+m} \omega(z)\, \mathrm{d} z\bigg|_{x=0,\; y=0}
\notag\\ &
=\frac{1}{j!}\frac{\partial^j}{\partial x^j}
\int_{-\infty}^{\infty}
\mathcal{K}_N(x,z)\, z^{m} \omega(z)\, \mathrm{d} z\bigg|_{x=0}
\notag\\ &
=\frac{1}{j!}\frac{\partial^j}{\partial x^j}\;x^m \bigg|_{x=0}
\notag\\ &
=\delta_{jm}\;.
\end{align}
In the case of matrix $A$ \eqref{Aphi}
the entries of the inverse matrix $A^{-1}$ are given by \eqref{inverse}
with the function $\mathcal{K}_N(x,y)$ build from formula \eqref{KN}
in terms of orthogonal polynomials $p_n(x)$, defined in Eq.~\eqref{PcalF}.

Generalizing the simple computation above one can evaluate the
entries of the product  matrix $A_{-}^{-1}A_{+}^{}$:
\begin{equation}\label{AA}
\left[A_{-}^{-1} A_{+}^{}\right]_{jm}
= \frac{1}{j!}\frac{\partial^{j}}{\partial x^j}
\int_{-\infty}^{\infty} \mathcal{K}^{-}_N (x,z)\, z^m
\,\omega^{+}(z)\,\mathrm{d} z\Bigg|_{x=0}.
\end{equation}
Here the $+$ or $-$ superscripts denote the dependence
on the variables $\phi_\pm$ respectively.
The determinant in \eqref{Zuqsl2} can now be transformed as follows
\begin{align}\label{trn}
&\ln {\det}_N (I- A_{-}^{-1}\, A_{+}^{})\,
\notag\\&\quad
=-\sum_{n=1}^{\infty}\frac{1}{n}
\tr_N^{}(A_{-}^{-1}A_{+}^{})^n
\notag\\&\quad
=-\sum_{n=1}^{\infty}\frac{1}{n}
\int_{-\infty}^{\infty}
\!\!\dots
\int_{-\infty}^{\infty}
\mathcal{K}^{-}_N(x_1,x_2)
\mathcal{K}^{-}_N(x_2,x_3)
\dots
\mathcal{K}^{-}_N(x_n,x_1)
\prod_{l=1}^n \omega^{+}(x_l) \mathrm{d} x_l
\notag\\&\quad
=: -\sum_{n=1}^{\infty}\frac{1}{n}
\tr(\mathcal{V}_N)^n
\notag\\[4pt]&\quad
=\ln {\det}(\mathcal{I}-\mathcal{V}_N)
\end{align}
where $\mathcal{V}_N$
is the integral
operator on the real axis with
kernel
\begin{align}\label{VN}
\mathcal{V}_N^{}(x,y)
=\mathcal{K}_N^{-}(x,y)\, \omega^{+}(y).
\end{align}
Using the  Christoffel-Darboux identity, this  kernel
 can be written as
\begin{equation}\label{VNI}
\mathcal{V}_N(x,y)
=\frac{\kappa_{N-1}^{-}}{\kappa_{N}^{-}}
\, \frac{p^{-}_N(x) p^{-}_{N-1}(y)
-p^{-}_{N-1}(x) p^{-}_N(y)}{x-y}\, \omega^{+}(y)
\end{equation}
rendering the integrability (in the sense of Ref. \cite{IIKS-90})
of integral operator $\mathcal{V}_N$ manifest.
Integral kernels of the form \eqref{VNI}
are also known under the name of correlation kernels since they arise
in expressions for eigenvalues correlation functions
in the theory of random matrices.
For discussion of the role and importance of this special
class of integral operators in connection with correlation functions
of integrable models and with the theory of random matrix,
see \cite{KBI-93,TW-94}.

Taking into account formulae \eqref{weight}, \eqref{Pcal},
\eqref{PcalF} and \eqref{kappa} we therefore get for the
partition function of the model,  with the choice
\eqref{uqsl2} for its   vertex weights,  the following
Fredholm determinant representation:
\begin{equation}%\label{}
{\tilde Z}_N
={\det}(\mathcal{I}-\mathcal{V}_N)
\end{equation}
where the kernel may be most explicitely written as
\begin{align}%\label{}
\mathcal{V}_N(x,y)
&=
\biggl\{
\hyper{-N}{\mathrm{i}x/2}{1}{1-\mathrm{e}^{-2\mathrm{i}\phi_{-}}}
\hyper{-N+1}{\mathrm{i}y/2}{1}{1-\mathrm{e}^{-2\mathrm{i}\phi_{-}}}
\notag\\ &\quad
-\hyper{-N+1}{\mathrm{i}x/2}{1}{1-\mathrm{e}^{-2\mathrm{i}\phi_{-}}}
\hyper{-N}{\mathrm{i}y/2}{1}{1-\mathrm{e}^{-2\mathrm{i}\phi_{-}}}
\biggr\}
\notag\\&\quad\times
\frac{N\mathrm{e}^{\mathrm{2iN}\phi_{-}}}{x-y}
\frac{\mathrm{e}^{\phi_{+}y}}{1-\mathrm{e}^{\pi y}+\mathrm{i} 0}\;.
\end{align}
This representation is valid for $0<\Re\phi_{+}<\pi$ and arbitrary
complex $\phi_{-}$.
Let us moreover underline that, contrarily to the original H\"ankel
determinant formula, the parameter $N$ can be extended here
from the set of positive
integers to the whole complex plane.

We shall now discuss alternative forms for  the representation we
have just obtained. First, it should be noted that
by shifting the integration contour and relabelling
the integration variables in each integral in \eqref{trn}
this result can also be put in the form
\begin{equation}\label{Zdisorder}
{\tilde Z}_N
={\det}\left(\mathcal{I}-\zeta\mathcal{W}_N\right)\;,\qquad
\zeta=\mathrm{e}^{\mathrm{i}(\phi_{-}-\phi_{+})}\;,
\end{equation}
with (to shorten the formulae we shall use more compact notations
in terms of polynomials)
\begin{align}\label{Vdisorder}
\mathcal{W}_N(x,y)
&=N\,\frac{
%\biggl\{
\MP{N}{1/2}{x}{\phi_{-}}
\MP{N-1}{1/2}{y}{\phi_{-}}
-\MP{N-1}{1/2}{x}{\phi_{-}}
\MP{N}{1/2}{y}{\phi_{-}}
}{x-y}
%\biggr\}
\notag\\ &\quad\times
\frac{\mathrm{e}^{2\phi_{+}y}}{1+\mathrm{e}^{2\pi y}}\;.
\end{align}
We see here that as a matter of fact the kernel is a real-valued
function for real $\phi_{\pm}$
(since Meixner-Pollaczek polynomials $\MP{n}{\lambda}{x}{\phi}$ are
real-valued for real $x$ and $\phi$).

The parametrization of
the weights in the form \eqref{|Delta|<1} with real $\lambda$ and $\eta$
(hence, real $\phi_{\pm}$)
is typical for the so-called disordered phase of the model \cite{B-82}.
Thus, the just obtained representation for the partition function,
even if valid for arbitrary choice of vertex weights,
can be regarded as ``adapted'' to the disordered phase.
The other two physical regimes are ferroelectric and
antiferroelectric and they can be obtained by choosing
$\phi_{\pm}$ to be purely imaginary.

To obtain corresponding representations it is sufficient to note that in the
case of complex $\phi_{+}$ the integrals in \eqref{trn} can be evaluated
by closing the integration contours upwards (downwards)
in the complex plane of the variables $x_1,\dots,x_n$
if $\Im\phi_{+}>0$ (if $\Im\phi_{+}<0$).
As a result, each integral is given
by a sum of residues at simple poles of the function $\omega(x)$
lying in the upper (respectively lower) complex half-plane.
Considering the case of $\Im\phi_{+}>0$ (the case of $\Im\phi<0$
leading  to an essentially equivalent result)
we obtain the following representation for the partition function:
\begin{equation}\label{Zferro}
{\tilde Z}_N
={\det} (\mathcal{I}-\widetilde{\mathcal{V}}_N)
\end{equation}
where $\widetilde{\mathcal{V}}_N$ is the integral operator
with discrete kernel
\begin{align}\label{Vferro}
\widetilde{\mathcal{V}}_N(x,y)
&=-\biggl\{
M_N\bigl(x;1,\mathrm{e}^{-2\tilde\phi_{-}}\bigr)
M_{N-1}\bigl(y;1,\mathrm{e}^{-2\tilde\phi_{-}}\bigr)
\notag\\ &\qquad
-M_{N-1}\bigl(x;1,\mathrm{e}^{-2\tilde\phi_{-}}\bigr)
M_N\bigl(y;1,\mathrm{e}^{-2\tilde\phi_{-}}\bigr)
\biggr\} \frac{N\mathrm{e}^{-2N\tilde\phi_{-}}}{x-y}\;
\mathrm{e}^{-2\tilde\phi_{+} y}
\end{align}
whose ``integration'' variables $x$, $y$ take nonnegative
integer values; for $x=y$ the kernel
is to be understood in the sense of the Cristoffel-Darboux
identity; in the last formula, the standard notation for Meixner
polynomials
\begin{equation}\label{MeixnerP}
M_n(x;\beta,c)=\hyper{-n}{-x}{\beta}{1-\frac{1}{c}}
\end{equation}
has been used \cite{KS-98}. The ``tilded'' variables
are defined as $\phi_\pm^{}=\mathrm{i}\tilde\phi_\pm\;$.
The representation \eqref{Zferro} is valid for $\Re\tilde\phi_{+}>0$
and arbitrary complex $\tilde\phi_{-}\;$.
The real values of $\tilde\phi_{\pm}$
correspond to ferroelectric ($\tilde\phi_\pm>0$) and
antiferroelectric ($\tilde\phi_{+}>0$, $\tilde\phi_{-}<0$)
phases of the model.

It is to be mentioned that representation \eqref{Zdisorder} can be also obtained
directly by employing the formula
\begin{equation}\label{coth}
\coth\tilde\phi=2\sum_{x=0}^{\infty} \mathrm{e}^{-2\tilde\phi x}\;,\qquad
\Re\tilde\phi>0
\end{equation}
instead of \eqref{cot} and repeating all considerations of the
previous Section. The appearance of Meixner polynomials
is then quite obvious since for $|c|<1$ they are subject to the
orthogonality condition
\begin{equation}\label{Moc}
\sum_{x=0}^{\infty} M_j(x;\beta,c) M_k(x;\beta,c)
\frac{(\beta)_x}{x!}\; c^x
= \frac{c^{-j}}{1-c}\,\delta_{jk}
\end{equation}
and upon setting $\beta=1$ and $c=\mathrm{e}^{-2\tilde\phi}$
the identification of the proper set of orthogonal polynomials
is achieved.

We conclude this Section by considering
the so-called rational parametrization of the vertex weights,
which  corresponds to the case  in which  $\mathsf{a}$, $\mathsf{b}$
and $\mathsf{c}$  are restricted
by the condition $\mathsf{a}\pm\mathsf{b}=\mathsf{c}\;$.
This regime can be obtained  through a suitable limit from
vertex weights  \eqref{|Delta|<1}.
Namely,  depending on the choice
of the sign in this restriction, one should just substitute
$\lambda,\eta\to\epsilon\lambda,\epsilon\eta$
(for plus sign)
or by $\lambda,\eta\to\pi/2-\epsilon\lambda,\pi/2-\epsilon\eta$
(for minus sign) in \eqref{|Delta|<1} and take the limit $\epsilon\to 0$,
after renormalization of the weights by a factor $1/\epsilon$. We should
of course recover in this limit the corresponding result of
Ref.~\cite{S-00}.

The H\"ankel determinant formula for
the partition function of the model with rational
weights is simply given by formula \eqref{Z=detZ} with
the rational functions $\lambda, \eta$
instead of sine functions:  $\sin(*)\to (*)$.
The Fredholm determinant representation for the partition
function in this case involves Laguerre polynomials;
the ``rational limit'' of the kernel
\eqref{Vdisorder} can be easily found using
\begin{equation}%\label{}
\lim_{\epsilon\to 0}\MP{n}{1/2}{x/\epsilon}{\epsilon\phi}=
L_n(-2 \phi x),\qquad
\lim_{\epsilon\to 0}
\frac{\mathrm{e}^{\epsilon\phi_{+}(y/\epsilon)}}{1+\mathrm{e}^{\pi
(y/\epsilon)}}
= \mathrm{e}^{\phi_{+} y} \theta(-y),
\end{equation}
where $L_n(x)$ is Laguerre polynomial,  and
$\theta(x)$ is Heaviside step-function.
The same result can be obtained from \eqref{Vferro}, too, with the
discrete measure  turning into the continuous one in the standard
way, when performing the rational limit. Explicitly, the partition
function in the rational case is given by the Fredholm determinant
of  the integral operator on the real positive half-axis, defined by
the kernel
\begin{equation}%\label{}
\mathcal{V_N}(x,y)=
-N\frac{L_{N}(\xi x)L_{N-1}(\xi y)-L_{N-1}(\xi x)L_{N}(\xi y)}{x-y}\;
\mathrm{e}^{-y}
\end{equation}
where $\xi=\phi_{-}/\phi_{+}$. The result of Ref.~\cite{S-00} is
thus reproduced in the rational limit, which corresponds there
to the case $q=1$.

%%%%%%%%%%%%%%%%%%%%%%%%%%%%%%%%%%%%%%%%%%%%%%%%%%%%%%%%%%%%%%%%%%
\section{The finite-size determinant representation}

The determinant in \eqref{Zuqsl2} can be also written
as that of some $N\times N$ symmetric matrix whose entries
are simply connected with the kernel of the integral operator
in the Fredholm determinant representation. One can insert \eqref{KN}
in the second line of \eqref{trn} and obtain
\begin{align}%\label{}
& \ln {\det}_N (I- A_{-}^{-1}\, A_{+}^{})\,
\notag\\&\quad
=-\sum_{n=1}^{\infty}\frac{1}{n}
\int_{-\infty}^{\infty}
\!\!\dots
\int_{-\infty}^{\infty}
\mathcal{K}^{-}_N(x_1,x_2)
\mathcal{K}^{-}_N(x_2,x_3)
\dots
\mathcal{K}^{-}_N(x_n,x_1)
\prod_{l=1}^n \omega^{+}(x_l) \mathrm{d} x_l
\notag\\&\quad
=-\sum_{n=1}^{\infty}\frac{1}{n}
\sum_{k_1,\dots,k_n=0}^{N-1}
V_{k_1k_2} V_{k_2k_3}
\dots
V_{k_nk_1}
\notag\\[4pt]&\quad
=\ln {\det}_N (I-V)
\end{align}
where entries of the matrix $V$ are
\begin{equation}%\label{}
V_{jk}^{}= \int_{-\infty}^{\infty}
p_j^{-}(x)p_k^{-}(x)\,\omega^{+}(x)\,\mathrm{d}x\;.
\end{equation}
Clearly, this identity is a consequence of the fact that the matrices
$A_{-}^{-1}A_{+}$ and $V$ are related by some similarity transformation.

Hence, the partition function admits also the representation
\begin{equation}\label{ZdetW}
{\tilde Z}_N
={\det}_N \bigl(I-V\bigr).
\end{equation}
It turns out that
in contrast to entries of matrix $H$,
entering the original H\"ankel determinant representation
\eqref{Z=detZ}, those of the matrix $V$ can be computed explicitly.
It is convenient (in analogy with \eqref{Zdisorder})
to introduce the matrix $W$ by
\begin{equation}%\label{}
V=\zeta W\;,\qquad
\zeta=\mathrm{e}^{\mathrm{i}(\phi_{-}-\phi_{+})}
=\mathrm{e}^{-2\mathrm{i}\eta}\;.
\end{equation}
Entries
of $W$ are defined by the formula
\begin{equation}\label{Wdisorder}
W_{jk}=2 \sin\phi_{-}
\int_{-\infty}^{\infty}
\MP{j}{1/2}{x}{\phi_{-}}
\MP{k}{1/2}{x}{\phi_{-}}
\frac{\mathrm{e}^{2x\phi_{+}}}{1+\mathrm{e}^{2\pi x}}\,\mathrm{d} x.
\end{equation}
As  shown in detail in the Appendix, the integral can be
evaluated in closed form. Introducing, for the sake of convenience,
the notations
\begin{equation}%\label{}
\beta=\frac{\sin\phi_{-}}{\sin\phi_{+}}
=\frac{\sin(\lambda-\eta)}{\sin(\lambda+\eta)}\;,\qquad
\gamma= \frac{\sin(\phi_{+}-\phi_{-})}{\sin\phi_{+}}
=\frac{\sin 2\eta}{\sin(\lambda+\eta)}\;,
\end{equation}
(note that $\beta$ and $\gamma$ are precisely the quantities entering
the $U_q(\mathfrak{sl}_2)$-invariant $R$-matrix, see eqs. \eqref{Rmatrix}
and \eqref{bgnu},
$\beta=\beta(\nu)$, $\gamma=\gamma(\nu)$)
the result reads
\begin{align}\label{Wmatrix}
W_{jk}
&=\gamma^{j+k+1}
\sum_{n=0}^{\min(j,k)} \binom{j}{n} \binom{k}{n}
\left(\frac{\beta}{\gamma}\right)^{2n+1}
\notag\\&=\beta \gamma^{j+k}
\hyper{-j}{-k}{1}{\frac{\beta^2}{\gamma^2}}\;.
\end{align}
It is worth to remind
that the representation \eqref{ZdetW} is for the partition
function with the weights \eqref{uqsl2}; the original normalization
can be achieved by multiplying the RHS of \eqref{ZdetW} by the factor
$[\sin(\lambda+\eta)]^{N^2} \mathrm{e}^{\mathrm{i}(\lambda-\eta)N}$.

The second expression in \eqref{Wmatrix} shows that
the entries of matrix $W$ are, in fact, Meixner polynomials,
see \eqref{MeixnerP}. On the other hand, from
the first expression in \eqref{Wmatrix} follows that
matrix $W$ can be written as a product of much simpler matrices.
Indeed, introducing $N\times N$ matrices
\begin{align} \label{JJJ}
(J_{+})_{nm}&=n\,\delta_{n-1,m}
\notag\\
(J_{0}\,)_{nm}&=(n+1/2)\,\delta_{n,m}
\notag\\
(J_{-})_{nm}&=(n+1)\,\delta_{n+1,m}
\end{align}
and taking into account that
\begin{equation}
\bigl(\exp\bigl\{\gamma J_{+}\bigr\}\bigr)_{nm}
=\gamma^{n-m}\binom{n}{m}\;,\qquad
J_{-}=(J_{+})^T\;,
\end{equation}
one obtains
\begin{equation} \label{Wgd}
W =\exp\bigl\{\gamma J_{+}\bigr\}\,
\beta^{\textstyle 2J_0}\,
\exp\bigl\{\gamma J_{-}\bigr\}\;.
\end{equation}
Thus, the matrix $W$ is defined by its Gauss decomposition,
i.e., as a product of a lower-triangular, a diagonal,
and an  upper-triangular matrices, respectively.

Matrix $W$ simplifies considerably in
the limit $N\to\infty$. This is a consequence of the fact that
the semi-infinite dimensional matrices
$J_{\pm,0}$ with entries \eqref{JJJ} satisfy $\mathfrak{su}(1,1)$ algebra
commutation relations
\begin{equation}\label{su11}
[J_{-},J_{+}] = 2J_{0}\;, \qquad [J_\pm,J_0]=\mp J_\pm\;.
\end{equation}
Standard techniques (widely used for instance in the
theory of generalized coherent states \cite{P-86})
can now be applied to \eqref{Wgd}, with the following result
\begin{equation}\label{Ninfty}
W = \exp\bigl\{2\eta K\bigr\}\qquad (N=\infty)\;,
\end{equation}
where the semi-infinite dimensional matrix $K$ is
\begin{equation}%\label{}
K = \frac{1}{\sin\nu}\,\bigl(J_{-}+J_{+}-2 \cos\nu\, J_0\bigr)\;.
\end{equation}
Here (and below) we use ``mixed'' set of variables,
$\nu=\phi_{-}=\lambda-\eta$ and $\eta$
(indeed the most natural ones for the
 $U_q(\mathfrak{sl}_2)$-invariant $R$-matrix, see
Eqs.~\eqref{Rmatrix}--\eqref{unitarity}).

The matrix $K$ is the  Jacobi matrix corresponding to the recurrence
relation of Meixner-Pollaczek, Laguerre or Meixner polynomials
depending whether $|\cos\nu|<1$,
$|\cos\nu|=1$ or $|\cos\nu|>1$,
respectively. These polynomials are therefore eigenfunctions
of $K$ with eigenvalue $x$,  and the spectrum of $K$ is
thus given by the support of the measure appearing in the
orthogonality condition of the corresponding polynomials,
which is $\mathbb{R}$, $\mathbb{R}_{+}$ or $\mathbb{Z}_{+}$, respectively
(i.e., it is continuous in the first and second case and discrete
in the third one). {}From this point of view the presence of these
same polynomials in the kernel of the integral operator in the Fredholm
determinant representations is obvious: the integral operator is
just the semi-infinite dimensional matrix $\Pi_N \exp\{2\eta K\}$
re-expressed in the basis of the eigenfunctions of $K$;
here the $N$-dimensional projector $\Pi_N$ is
 the matrix with its first $N$ diagonal entries equal to one,
and  all others  entries equal to zero.

Another point which is to be discussed is that formula
\eqref{Ninfty} tells us that in case $N=\infty$ , matrix $W$ (and hence $V$)
is just the exponential of some ``simple'' matrix.
The analogy which come  into mind after looking at formulae
\eqref{ZdetW} and \eqref{Ninfty} is with a typical result of
calculation of traces of certain class of operators acting over
the  Fock space of $N$ fermions. Namely, given  a matrix $A$,
and the corresponding quantum operator $\op{A}$, bilinear
in canonical fermion operators, built from matrix $A$ as follows
\begin{equation}\label{opA}
\op{A}= \sum_{n,m=0}^{N-1}
\hat c_{n+1}^\dagger A_{nm}^{}\,\hat c_{m+1}^{},\qquad
\hat c_n^\dagger \hat c_m^{} + \hat c_n^{} \hat c_m^\dagger = \delta_{nm}
\end{equation}
it is well-known \cite{B-66}  that
\begin{equation}\label{Str}
\Tr\bigl[\exp \op{A}\bigr]={\det}_N\bigl(I+\exp A\bigr)\;.
\end{equation}
To make a connection with formula \eqref{ZdetW},
it is to be mentioned that the minus sign
in RHS in \eqref{Str} can be acquired
by considering  supertraces instead of traces;
the supertrace is defined as
$\Str\bigl[\,*\,\bigr]:=\Tr\bigl[(-1)^\op{N}\,*\,\bigr]$,
where $\op{N}$ denotes the fermion number operator:
$\op{N}=\sum_{n=1}^{N}\hat c_n^\dagger \hat c_n^{}\;$.
However, since
the fermion number operator commutes with any operator of the form
\eqref{opA}, one can consider just traces when dealing with the
exponentials of such operators.
Taking into account this we can therefore
write the following trace formula for the partition function
\begin{equation}%\label{}
{\tilde Z}_N
=\Tr\left[\exp\bigl\{2\eta \op{L}_N\bigr\}\right]
\end{equation}
with
\begin{equation}%\label{}
\op{L}_N= \op{K}_N + \mathrm{i}\frac{\pi-2\eta}{2\eta}\, \op{N}\;,
\end{equation}
where the operator $\op{K}_N$ is built from the matrix
\begin{equation}
K_N = \frac{1}{2\eta}
\ln W\;.
\end{equation}
Note that in contrast to the entries of matrix $W$ those of
matrix $K_N$ (and hence of $L_N$) essentially depend on $N$ as
indicated by the notations. Operator $\op{K}_N$,
though bilinear in fermions, is essentially
non-local, thus making the entries of matrix $K_N$
for finite $N$ rather complicated quantities. However,
formula \eqref{Ninfty} implies that matrix $K_N$ simplifies considerably
in the case $N=\infty$, and the corresponding
operator $\op{K}:=\op{K}_\infty$ becomes a local one.
Explicitly we obtain
\begin{equation}\label{boost}
\op{K} = \sum_{n=1}^{\infty} n\, \op{H}_{n,n+1}\;,
\end{equation}
where
\begin{equation}%\label{}
\op{H}_{n,n+1} = \frac{1}{\sin\nu}\left\{
\hat c_n^\dagger \hat c_{n+1}^{}
+\hat c_{n+1}^\dagger \hat c_n^{}
-\cos\nu\,
\bigl(\hat c_n^\dagger \hat c_n^{}
+ \hat c_{n+1}^\dagger \hat c_{n+1}^{}\bigr)
\right\}\;.
\end{equation}
We recognize in this expression for operator $\op{K}$
what is known in the literature
as a boost,  or  ladder,  operator. To be precise
$\op{K}$ is the positive half-axis part $\op{K}=\op{B}^{(+)}$
of the total boost operator $\op{B}=\op{B}^{(-)}+\op{B}^{(+)}$,
with $[\op{B}^{(+)},\op{B}^{(-)}]=0$, see \cite{Th-98}.
Operator $\op{B}$ is the
boost operator for the model described by the following Hamiltonian
\begin{equation}%\label{}
\op{H}=\sum_{n}^{} \op{H}_{n,n+1}=
\frac{1}{\sin\nu} \op{H}_0 - 2 \cot\nu \op{N}
\end{equation}
where $\op{H}_0$ is the hopping term. From this formula it is
clear that $\nu$,  the spectral parameter in the six-vertex model,
is here playing the role of a chemical potential rather than a
coupling constant. Nevertheless, all the construction above has
led us to a description which is quite analogous to what we have
in the corner transfer matrix formalism \cite{B-82}, or to a more
general extent, in the vertex operator approach to integrable
models \cite{JM-94} and angular quantization method in integrable
quantum field theory (see, e.g., \cite{KCP-99} an references
therein).

As previously explained, in the case of finite $N$ operator
$\op{K}_N$ is nonlocal (though free-fermionic). One can however
still write the partition function in terms of the product of exponentials
of local operators.
As a generalization of \eqref{Str} the following is also valid
\begin{equation}\label{Tr-prod}
\Tr\biggl[\prod_i \exp \op{A}_i\biggr]
={\det}_N\biggl(I+\prod_i \exp A_i\biggr)
\end{equation}
where operators $\op{A}_i$ are constructed out of matrices
$A_i$ through formula \eqref{opA}. We can therefore
write the following ``reconstruction''  trace formula for the
partition function
\begin{equation}\label{Ztr}
{\tilde Z}_N
=\Tr\left[(-\mathrm{e}^{-2\mathrm{i}\eta})^{\textstyle\op{N}}
\exp\bigl\{\gamma\op{J}_{+}\bigr\}\,\beta^{\textstyle 2\op{J}_0}\,
\exp\bigl\{\gamma \op{J}_{-}\bigr\}
\right]
\end{equation}
with
\begin{equation}%\label{}
\op{J}_{+}=\sum_{n=1}^{N-1} n\, c_{n+1}^\dagger c_{n}^{}\;,\qquad
\op{J}_0=\sum_{n=1}^{N} (n-1/2)\, c_n^\dagger c_n^{}\;,\qquad
\op{J}_{-}=(\op{J}_{+})^\dagger\;.
\end{equation}
The reconstruction formula \eqref{Ztr} expresses the partition function
of the six-vertex model as the trace of a quantum operator
built out of fermions, or, modulo Jordan-Wigner transformation,
of spin-$1/2$ operators.
Note that, in fact, we have derived these expressions
for the partition function starting from the
H\"ankel determinant representation,
previously obtained within the Quantum Inverse Scattering Method.
The question which arises from the considerations presented
here, is how representation
\eqref{Ztr} in terms of fermions  (or spins),
which are to be  interpreted  as effective degrees of freedom,
could be extracted directly from the basic definition of the model
in terms of vertex configurations.
An answer to this question might suggest alternative
approaches to  the  open problem of calculation of
correlation functions for the model.

%%%%%%%%%%%%%%%%%%%%%%%%%%%%%%%%%%%%%%%%%%%%%%%%%%%%%%%%%%%%%%%%%%%
\section*{Acknowledgments}

We are grateful to N.M. Bogoliubov,  A. Cappelli and
V. Tognetti for useful discussions. We acknowledge
financial support from MIUR COFIN programme  and from
INFN (Iniziativa Specifica FI11). This work was been partially
done within the European
Community network ``EUCLID'' (HPRN-CT-2002-00325).
One of us (A.P.) is supported in part by Russian Foundation for Basic
Research, under Grant No. 01-01-01045, and by the programme ``Mathematical
Methods in Nonlinear
Dynamics'' of Russian Academy of Sciences.

F.C. thanks  Euler
International Mathematical Institute  and Steklov Institute of
Mathematics at St. Petersburg for warm hospitality.
A.P. is very grateful to Florence Department of I.N.F.N. and
Physics Department of the University of Florence
for kind hospitality and support that make
this collaboration possible.

%%%%%%%%%%%%%%%%%%%%%%%%%%%%%%%%%%%%%%%%%%%%%%%%%%%%%%%%%%%%%%%%%%%
\appendix
\renewcommand{\theequation}{A.\arabic{equation}}
\setcounter{equation}{0}
\section*{Appendix}

The integral defining the entries of matrix $W$ in \eqref{Wdisorder}
is a particular case of the integral
\begin{equation}\label{int}
I_{nm}^{(\lambda)}(\tau,\omega;\phi)
=\frac{1}{2\pi}\int_{-\infty}^{\infty}
\MP{n}{\lambda}{x}{\tau}
\MP{m}{\lambda}{x}{\omega}
|\Gamma(\lambda+\mathrm{i} x)|^2
\mathrm{e}^{(2\phi-\pi) x}\,\mathrm{d} x
\end{equation}
where $\phi\in(0,\pi)$ and $\lambda$ is assumed to
be real and positive, $\lambda>0$.
These restriction are important for convergence of the integral.
Our aim here is to prove that for arbitrary $\tau$ and $\omega$
the quantity $I_{nm}^{(\lambda)}(\tau,\omega;\phi)$ has the following
expression
\begin{align}\label{Inm}
I_{nm}^{(\lambda)}(\tau,\omega;\phi)
&= \frac{\Gamma(2\lambda+n)\,\Gamma(2\lambda+m)}
{(2\sin\phi)^{2\lambda}\,\Gamma(2\lambda)\, n!\, m!}
\left[\frac{\sin(\tau-\phi)}{\sin\phi}\right]^n
\left[\frac{\sin(\omega-\phi)}{\sin\phi}\right]^m
\notag\\ &\quad \times
\hyper{-n}{-m}{2\lambda}{\frac{\sin\tau\sin\omega}
{\sin(\tau-\phi)\sin(\omega-\phi)}}.
\end{align}
Equality \eqref{Inm} is a consequence of the orthogonality condition
for the Meixner-Pollaczek polynomials and of the identity
\begin{align}\label{MP=MP}
\MP{n}{\lambda}{x}{\tau} =
\sum_{k=0}^{n}
\frac{\Gamma(n+2\lambda)}{\Gamma(k+2\lambda)\, (n-k)!}\;
\frac{\bigl[\sin(\tau-\phi)\bigr]^{n-k}
\bigl(\sin\tau\bigr)^k}{\bigl(\sin\phi\bigr)^n}\;
\MP{k}{\lambda}{x}{\phi}\;,
\end{align}
which can be viewed as the extension of an analogous formula for Laguerre
polynomials (see,  e.g.,  Ref. \cite{E-81}, \S 10.12, Eqn. (40)).
Indeed, using identity
\eqref{MP=MP} for both polynomials under the integration  sign and employing
the orthogonality condition \eqref{MPoc} one immediately obtains
\begin{align}%\label{}
&I_{nm}^{(\lambda)}(\tau,\omega;\phi)
= \frac{\bigl[\sin(\tau-\phi)\bigr]^n
\bigl[\sin(\omega-\phi)\bigr]^m
}{2^{2\lambda}(\sin\phi)^{n+m+2\lambda}}
\notag\\ &\qquad\times
\sum_{k=0}^{\min(n,m)}\!\!
\frac{\Gamma(n+2\lambda)\,\Gamma(m+2\lambda)}
{\Gamma(k+2\lambda)\, (n-k)!\, (m-k)!\, k!}
%\notag\\ &\quad\times
\left[\frac{\sin\tau \sin\omega}{\sin(\tau-\phi)\sin(\omega-\phi)}
\right]^k\;.
\end{align}
Rewriting the finite sum here as a truncated
hypergeometric series results in expression \eqref{Inm}.

Let us prove now the identity \eqref{MP=MP}.
To simplify as much as possible the combinatorics,
let us consider the three-term  relation satisfied by the Meixner-Pollaczek
polynomials:
\begin{align}\label{MPrr}
&
(n+1)\,\MP{n+1}{\lambda}{x}{\phi}-
\bigl[2x \sin\phi + 2(n+\lambda)\cos\phi\bigl]\MP{n}{\lambda}{x}{\phi}
\notag\\ &\qquad
+ (n+2\lambda-1)\, \MP{n-1}{\lambda}{x}{\phi} =0\;.
\end{align}
Let us define
\begin{equation} \label{Sn}
S_n(x;\phi)=(\sin\phi)^{-n}
\MP{n}{\lambda}{x}{\phi}\;.
\end{equation}
Recurrence relation \eqref{MPrr} takes the form
\begin{align}\label{Snxf}
&
(n+1)\;S_{n+1}(x;\phi)
- 2(n+\lambda)\cot\phi\;S_n(x;\phi)
\notag\\ &\qquad
+ (n+2\lambda-1)\bigl[1+(\cot\phi)^2\bigr]\; S_{n-1}(x;\phi)
=2x\,S_n(x;\phi)\;.
\end{align}
{}From this relation it is clear that $S_n(x;\phi)$ depends on $\phi$ only
through $\cot\phi$, and, moreover, it is a polynomial of order $n$
in $\cot\phi$. It is useful to consider the case $\phi=\pi/2$
so that $\cot\phi=0$. Denoting $S_n(x) =S_n(x;\pi/2)$,
in this case one has
\begin{gather}\label{Snx}
(n+1)\,S_{n+1}(x)
+ (n+2\lambda-1) S_{n-1}(x)=2x S_n(x).
\end{gather}
Let us consider the semi-infinite dimensional matrices
\begin{align}\label{Jpmz}
\bigl(J_{-}\bigr)_{nm} &= (n+1)\, \delta_{n+1,m}
\notag\\
\bigl(J_{0}\,\bigr)_{nm} &= (n+\lambda)\, \delta_{n,m}
\notag\\
\bigl(J_{+}\bigr)_{nm} &= (n+2\lambda-1)\, \delta_{n-1,m}
\end{align}
where, as in the main text of the paper, $n,m\in \{0,1,2,\dots\}$.
These matrices obey $\mathfrak{su}(1,1)$ algebra commutation relations
\eqref{su11}. Modulo  a diagonal similarity transformation,
they correspond to the standard matrix realization of the positive
discrete representation $\mathcal{D}^{(+)}(\lambda)$
of $\mathfrak{su}(1,1)$. In terms of matrices \eqref{Jpmz}
relations \eqref{Snxf} and \eqref{Snx} read
\begin{align}%\label{}
&\bigl[
J_{-} - 2\cot\phi\, J_0 +  \bigl(1+(\cot\phi)^2\bigr) J_{+}
\bigr]\, \vec S(x;\phi)= 2x\, \vec S(x;\phi)
\notag\\
&
\bigl[ J_{-} +  J_{+} \bigr]\, \vec S(x)= 2x\, \vec S(x)\;,
\end{align}
where $n$-th  component of $\vec S(x,\phi)$ is just $S_n(x,\phi)$,
and $\vec S(x)\equiv \vec S(x,\pi/2)$.
Using the commutation relations of $\mathfrak{su}(1,1)$ algebra
it can be straightforwardly checked that the following relation
is valid
\begin{equation}%\label{}
\exp\{\alpha J_{+}\}
\bigl[ J_{-} + J_{+} \bigr]
=\bigl[ J_{-} -2\alpha J_0+ \bigl(1+\alpha^2\bigr)J_{+}\bigr]\,
\exp\{\alpha J_{+}\}.
\end{equation}
Thus, by identifying $\alpha=\cot\phi$ we obtain
\begin{equation}\label{key}
\vec S (x;\phi) =\exp\{\cot\phi\, J_{+}\}\, \vec S (x)\;.
\end{equation}
This is a key identity to prove relation \eqref{MP=MP}.
Indeed relation \eqref{key} implies that
\begin{equation}%\label{}
\vec S (x;\tau) =\exp\{(\cot\tau-\cot\phi)J_{+}\}\, \vec S (x;\phi)\;.
\end{equation}
Then, taking into account that
\begin{equation}%\label{}
\bigl(\exp\{\alpha J_{+}\}\bigr)_{nm}
=\frac{\Gamma(n+2\lambda)}{\Gamma(m+2\lambda)\, (n-m)!}\;
\alpha^{n-m}
\end{equation}
one obtains
\begin{equation}%\label{}
S_n(x;\tau) = \sum_{m=0}^{n}
\frac{\Gamma(n+2\lambda)}{\Gamma(m+2\lambda)\, (n-m)!}\;
(\cot\tau-\cot\phi)^{n-m} S_m(x;\phi).
\end{equation}
Rewriting the last equation in terms of the standard Meixner-Pollaczek
polynomials, see \eqref{Sn}, one arrives finally to identity \eqref{MP=MP},
which is thus proved.

%%%%%%%%%%%%%%%%%%%%%%%%%%%%%%%%%%%%%%%%%%%%%%%%%%%%%%%%%%%%%%%%%%%

\end{document}